# Observation of intrinsically bright terrestrial gamma ray flashes from the Mediterranean basin

T. Gjesteland[1,2], N. Østgaard[2], S. Laviola[3], M. M. Miglietta[4], E. Arnone[3], M. Marisaldi[2,5], F. Fuschino[5], A. B. Collier[6], F. Fabró[7], and J. Montanya[7]

[1]Department of Engineering Sciences, University of Agder, Grimstad, Norway, [2]Birkeland Centre for Space Science, Department of Physics and Technology, University of Bergen, Bergen, Norway, [3]CNR-ISAC, Bologna, Italy, [4]CNR-ISAC, Lecce, Italy, [5]INAF-IASF Bologna, Bologna, Italy, [6]School of Chemistry and Physics, University of KwaZulu-Natal, Durban, South Africa, [7]Department of Electrical Engineering, Polytechnical University of Catalonia, Barcelona, Spain

**Abstract** We present three terrestrial gamma ray flashes (TGFs) observed over the Mediterranean basin by the Reuven Ramaty High Energy Solar Spectroscope Imager (RHESSI) satellite. Since the occurrence of these events in the Mediterranean region is quite rare, the characterization of the events was optimized by combining different approaches in order to better define the cloud of origin. The TGFs on 7 November 2004 and 16 October 2006 came from clouds with cloud top higher than 10–12 km where often a strong penetration into the stratosphere is found. This kind of cloud is usually associated with heavy precipitation and intense lightning activity. Nevertheless, the analysis of the cloud type based on satellite retrievals shows that the TGF on 27 May 2004 was produced by an unusual shallow convection. This result appears to be supported by the model simulation of the particle distribution and phase in the upper troposphere. The TGF on 7 November 2004 is among the brightest ever measured by RHESSI. The analysis of the energy spectrum of this event is consistent with a production altitude $\leq 12$ km, which is in the upper part of the cloud, as found by the meteorological analysis of the TGF-producing thunderstorm. This event must be unusually bright at the source in order to produce such a strong signal in RHESSI. We estimate that this TGF must contain $\sim 3 \times 10^{18}$ initial photons with energy >1 MeV. This is 1 order of magnitude brighter than earlier estimations of an average RHESSI TGF.

## 1. Introduction

Terrestrial gamma ray flashes are bursts of gamma radiation originating from thunderclouds and were first discovered by the Compton Gamma Ray Observatory in 1994 [*Fishman et al.*, 1994]. TGFs have a typical duration of a few hundred microseconds and contain photons with energy up to several tens of MeV [*Smith et al.*, 2005; *Marisaldi et al.*, 2010]. Due to the high atmospheric absorption, TGFs were first claimed to be produced at altitudes > 30 km in order to escape to satellite altitude [*Fishman et al.*, 1994]. However, detailed analysis of the Reuven Ramaty High Energy Solar Spectroscopic Imager (RHESSI) data showed that the cumulative TGF energy spectrum was consistent with production at ~15 km altitude [*Dwyer and Smith*, 2005]. Other studies of gamma attenuation [*Williams et al.*, 2006] and spectral shape [*Carlson et al.*, 2007; *Østgaard et al.*, 2008; *Gjesteland et al.*, 2010] have confirmed this result, and it is now widely accepted that TGFs are produced via bremsstrahlung from high-energy electrons that are accelerated in electrical fields inside thunderstorms [*Dwyer et al.*, 2012]. Most TGF observations are obtained from the tropical region, where lightning occurrence is also the highest [*Smith et al.*, 2010; *Fabro et al.*, 2015]. However, there are also examples of TGFs in subtropical regions [*Grefenstette et al.*, 2009; *Gjesteland et al.*, 2012] . In this paper we present three TGFs from the Mediterranean basin. Two of the TGFs (27 May 2004 and 16 October 2006) are new events identified by the search algorithm presented by *Gjesteland et al.* [2012] and one of the events (7 November 2004) was previously presented by [*Smith et al.*, 2007] as an unusual TGF. These TGFs occurred in the subtropical region, which is outside of the typical TGF-producing region. For each of the three events we have carefully analyzed the meteorological properties of the TGF-producing thunderstorm. For the bright TGF on 7 November 2004 we have preformed a spectral analysis. For the TGF on 16 October 2006 we present lightning location data. For all three TGFs we have estimated the initial brightness of these TGFs, their peak current moment, and total charge moment change.

## 2. Data and Observation Methods

The TGFs studied in this paper were measured by RHESSI, which is a small NASA explorer designed to study solar flares. The detector will detect omnidirectional gamma fluxes and will therefore detect TGFs [*Smith et al.*, 2005; *Grefenstette et al.*, 2009; *Gjesteland et al.*, 2012]. The RHESSI instrument consists of nine Germanium detectors inside an aluminum cryostat. The energy range used for TGF detection is 25 keV to 17 MeV [*Grefenstette et al.*, 2009]. Due to the aluminum cryostat covering the RHESSI detectors, low-energy photons may be absorbed or scattered. The measured energy of the low-energy counts is therefore uncertain. The total effective area of the RHESSI detector is 239 cm$^2$ for a typical TGF spectrum [*Grefenstette et al.*, 2008]. The RHESSI relative time resolution is 1 binary microsecond ($2^{-20}$ s).

The three TGFs analyzed in this study were identified by the search algorithm presented by *Gjesteland et al.* [2012]. Two of the events were too dim to meet the requirements of the first RHESSI catalog. The last event has been previously reported as an unusually bright TGF [*Smith et al.*, 2007].

The convection of the TGF-producing thunderstorm is studied by the MicroWave Cloud Classification (MWCC) method [*Miglietta et al.*, 2013]. The computational core of the MWCC algorithm exploits the three water vapor absorption frequencies at 183.31 $\pm$ 1/ $\pm$ 3/ $\pm$ 7 GHz on board the Advanced Microwave Sounding Unit-B/Microwave Humidity Sounder flying on the NOAA and MetOp satellites [*Funatsu et al.*, 2007; *Hong et al.*, 2005]. Due to the vertical development of the different cloud types, the typical extinction of radiation at 183.3 $\pm$ 1/ $\pm$ 3/ $\pm$ 7 GHz in clear-sky conditions is perturbed as a function of cloud type and cloud vertical development. Stratified thin clouds, for example, usually have less impact on the water vapor channels peaking at lower altitudes and often appear transparent or completely masked by the absorption overlaying atmospheric water vapor. On the contrary, thick stratus clouds or convective cells, due to their pronounced vertical development, significantly perturb the radiation path. Therefore, from the analysis of the signal variations in the channels at 183.3 $\pm$ 1/ $\pm$ 3/ $\pm$ 7 GHz, which peak at different altitudes in the atmosphere, it is possible to detect the presence of clouds by assessing their altitude and type, either convective or stratiform. Recently, the computational scheme of the MWCC was improved with a probability-based module for hail detection and is currently in testing stage. In support to the MWCC cloud type/altitude estimates a robust test (hereafter Brightness temperature difference (BTD)) based on the combination of the Meteosat Second Generation (MSG) Spinning Enhanced Visible and Infrared Imager channels at 6.2μm and 10.8μm has been applied to detect the presence of the overshooting top formed during deep convection [*Schmetz et al.*, 1997]. Finally, the results from this method were compared with the output of the Weather Research and Forecasting (WRF) model giving an indication of the cloud distribution in the upper atmosphere. The Advanced Research Weather Research and Forecasting (WRF) model version 3.1 (see www.wrf-model.org) [*Skamarock et al.*, 2008] is used in the present study. It is a numerical weather prediction system that solves the fully compressible, nonhydrostatic Euler equations using a terrain-following hydrostatic-pressure vertical coordinate. In the present simulations 41 vertical levels are employed, more closely spaced in the planetary boundary layer. Simulations are performed on two domains, with a grid spacing of 16 km and 4 km, extending, respectively, 108 × 108 and 145 (in eastwest) × 169 grid points (in north-south), one-way nested one into the other. The domains are centered in the area where convection was observed, and the simulations last 36 h, covering the period where TGFs were reported. European Centre for Medium-Range Weather Forecasts analysis is used as initial and boundary conditions (updated every 6 h). The following parameterization schemes were used: the *Thompson et al.* [2004] microphysics; the *Kain* [2004] cumulus parameterization in the coarser grid (no parameterization is used in the inner grid); the rapid radiative transfer model for long-wave radiation, based on *Mlawer et al.* [1997]; the *Dudhia* [1989] scheme for shortwave radiation; the Mellor Yamada Janjic, a turbulent kinetic energy closure scheme for the boundary layer [*Janjic*, 2002]; and the Noah land surface model [*Niu et al.*, 2011].

For the TGF on 16 October 2006 we also have lightning data. The World Wide Lightning Location Network (WWLLN) is a global lightning network that measures lightning with temporal accuracy ∼30 μs and spatial accuracy of < 10 km [*Rodger et al.*, 2005; *Jacobson et al.*, 2006]. In 2006, which is the time of interest, WWLLN had a detection efficiency of ∼5 to 6% for all lightning strokes and ∼15% for cloud to ground (CG) strokes [*Rodger et al.*, 2009]. For 2006 WWLLN had a 12.4 % match with RHESSI TGFs [*Collier et al.*, 2011; *Gjesteland et al.*, 2012].

LINET is a 3D lightning detection network in Europe developed at the University of Munich, and started its operation 1 May 2006 [*Betz et al.*, 2009]. LINET can detect both intracloud (IC) and CG lightning. For lightning occurring with a baseline less than ∼ 250 km LINET is capable of distinguishing between CG and IC and it also gives IC emission altitudes. LINET also provides lighting stroke peak current estimates.

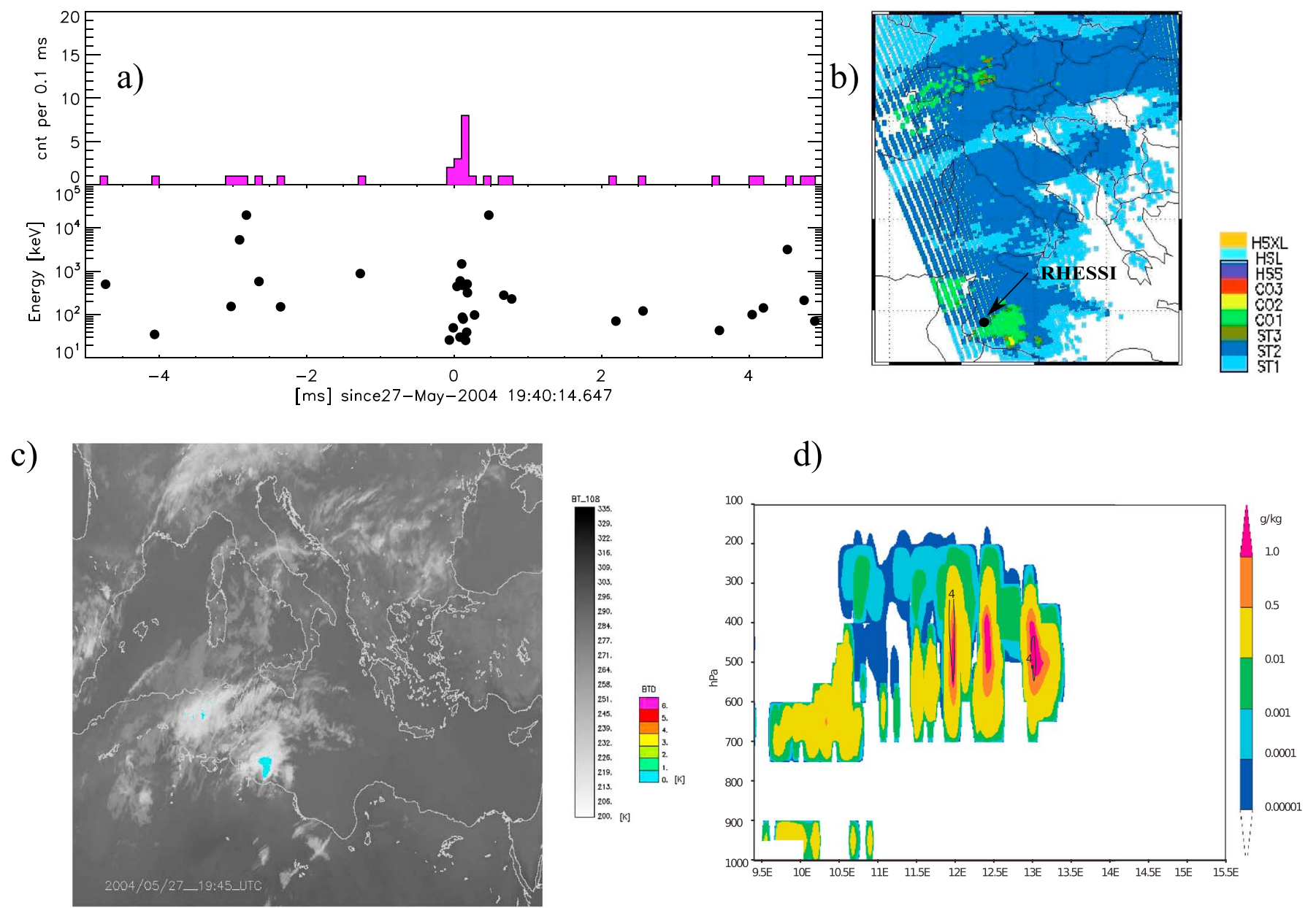

**Figure 1.** (a) The lightcurve and a scatterplot of photon energy and time for the RHESSI TGFs on 27 May 2004. (b) MWCC at 20:32 UTC. A weak convection (C01) is formed close to the TGF measurements. (c) MSG 10.8 μm cloud temperature measured at 19:45 in grey scale. BTD is shown in color. The maps range is latitude 25° to 50° and longitude −5° to 40°. (d) WRF model inner grid output: cloud water and cloud ice content (colors) and vertical velocity (contours) along the cross section where the convection is more developed top, cross section at 33.75°N, at 2100 UT. For the vertical velocity, only the values $w = 4, 8, 12$ m/s are shown.

CESI - SIRF (Centro Elettrotecnico Sperimentale Italiano - Sistema Italiano Rilevamento Fulmini) is an Italian lightning detection network based on Vaisala technology [*Cummins et al.*, 1998] which started its operation in 1994 [*Iorio and Ferrari*, 1996] and detects CG lightning over the Italian territory. It provides location, polarity type of lightning, and peak current estimates.

## 3. Observations

### 3.1. The 27 May 2004 Event

This TGF occurred in a well-organized convection close to the coastline of Tunis. Figure 1a shows the lightcurve of the TGF measured by RHESSI (top) and a scatterplot of time and energy for each detected photon (bottom). This is a weak TGF containing only 13 counts, and it has a duration of 0.312 ms. The duration is calculated by taking $\pm 2\sigma$ of a Gaussian fit to the lightcurve. Figure 1b shows a map with RHESSI location at the time of the event, and the convective clouds are shown in color. The convective cloud type is determined by the MWCC method. The MWCC measurements were obtained at 20:34 UTC, which is 54 min after the TGF measurement. The convection cloud type is classified as weak convection with cloud top height (−60°C) in the middle troposphere (around 6 km). Although the sensitivity of the high-frequency microwaves is linked to the scattering by large particles, the results of the MWCC are supported by the independent BTD test searching the possible intrusion into the stratosphere due to extreme thunderstorms. The BTD test based on data from 20:30 UT demonstrates that a very small area (169 MSG pixels) shoots out of the convection up to the stratosphere indicating that it was not an intense thunderstorm. The BTD test based on data from 19:45 UT (blue pixels in Figure 1c) shows similar behavior. To further investigate this case, a simulation with the WRF model shows that relatively shallow convection developed during the event. The model runs show peculiar conditions since there is a strong inhibition extending along an unusually deep layer (about 3000 m above the ground), due to the presence of very dry air of continental origin in the low levels and convective clouds extending above from about 650 g/cm$^2$. Compared with the other cases, the cloud extension, considering the volume within the cloud contour lines of 0.5 g/kg, remained confined well below the tropopause height (which is at about 250 g/cm$^2$), and the larger content was confined below 400 g/cm$^2$, probably as a consequence of the weak vertical motion (peak of about 5 m/s) confined to lower levels. (Below 300 g/cm$^2$; Figure 1d).

In agreement with this result, the atmosphere temperature profile measured at Trapani Radiosonde Station 18:00 UTC shows a minimum atmospheric temperature of 217 K between 10 and 11 km altitude. MSG measured an average cloud top temperature of 220–225 K which corresponds to an altitude of 10 km (316 g/cm$^2$).

### 3.2. The 7 November 2004 Event

This TGF occurred in the open Mediterranean basin. The TGF came from a convective system where two deep cores were formed. The left part of Figure 2a shows the lightcurve and a scatterplot of the TGF measured by RHESSI. This TGF is among the brightest TGFs measured by RHESSI. The TGF contains 79 counts and has a duration of 0.816 ms. Due to the high counting rate several photons are lost in the read out electronics for this event. This is called dead time losses. If we use the method presented in Østgaard et al. [2012] to determine total number of photons we estimate that 115 ± 15 photons hit the RHESSI detectors for this TGF.

Figure 2b shows RHESSI location at the time of the measurements and the convective clouds. The RHESSI location was between two convective cores. The MWCC results are at 16:15 UTC, 48 min before the TGF and the convection type classification is C03 indicating strong convection in the upper atmosphere. Furthermore, few pixels in the MWCC retrieval are associated with hail particles or super cooled water overlaying the ice particle layers [Muller et al., 1994], which sometimes express a radiometrically similar signature in passive microwaves. The confirmation of the frozen top is issued by the BTD test based on data from 16:15 UT describing a compact overshooting top (1843 MSG pixels) overlying the convective cores marking the presence of intrusion into the stratosphere consistent with MWCC results. In this case, the modeled cloud tops reach the level of 200 g/cm$^2$; thus, they cross the tropopause, which the model simulates at about 250 g/cm$^2$. The instability was moderate (Convective Available Potential Energy (CAPE) of about 1200 J/kg), while the inhibition was weak, and the ascending motion was relatively strong (about 9 m/s) even at 300 g/cm$^2$. The overshooting top is also seen in MSG at 17:00 UT (Figure 2c). The average cloud top temperature was 210–213 K which corresponds to an altitude of 12 km (227 g/cm$^2$).

### 3.3. The 16 October 2006 Event

The TGF discussed in this section was generated by a typical mesoscale convective system in the Mediterranean basin. The MWCC shows that the most active convective cores had two contributions (Figure 3b). The smallest was located in the western part close to the Greek coastline while the biggest, which produced the TGF, developed near the west coast of Sicily. Both systems are characterized by deep convection in the upper atmosphere with the top altitude higher than 8–10 km. Furthermore, a dark area corresponding to deep convection shows the presence of high scattering by ice typically associated with graupel particles. The BTD test based on data from 05:00 UT corroborates the MWCC results showing a dense distribution of cold particles penetrating into the stratosphere and forming a wide overshooting top. Figure 3b shows the RHESSI location and the convective clouds. The MWCC measurements are at 04:54 UTC, which is 2.5 h before the TGF.

For this case, the WRF model simulated unusually deep convective clouds. The cloud tops (defined as in section 3.1) reached 150 g/cm$^2$; thus, they clearly cross the tropopause, which in the present case can be estimated at about 250 g/cm$^2$. The instability for this case was very strong (simulated Convective Available Potential Energy (CAPE) of 2500 J/kg), while the inhibition was negligible. It is relevant that in this case an intense vertical motion affects the upper troposphere, with vertical velocity still larger than 12 m/s at about 250 g/cm$^2$. The BTD test based on data from 07:15 UT is shown in Figure 3c. The average MSG cloud top temperature was 205–210 K which corresponds to an altitude $\geq$ 12 km ($\leq$ 234 g/cm$^2$).

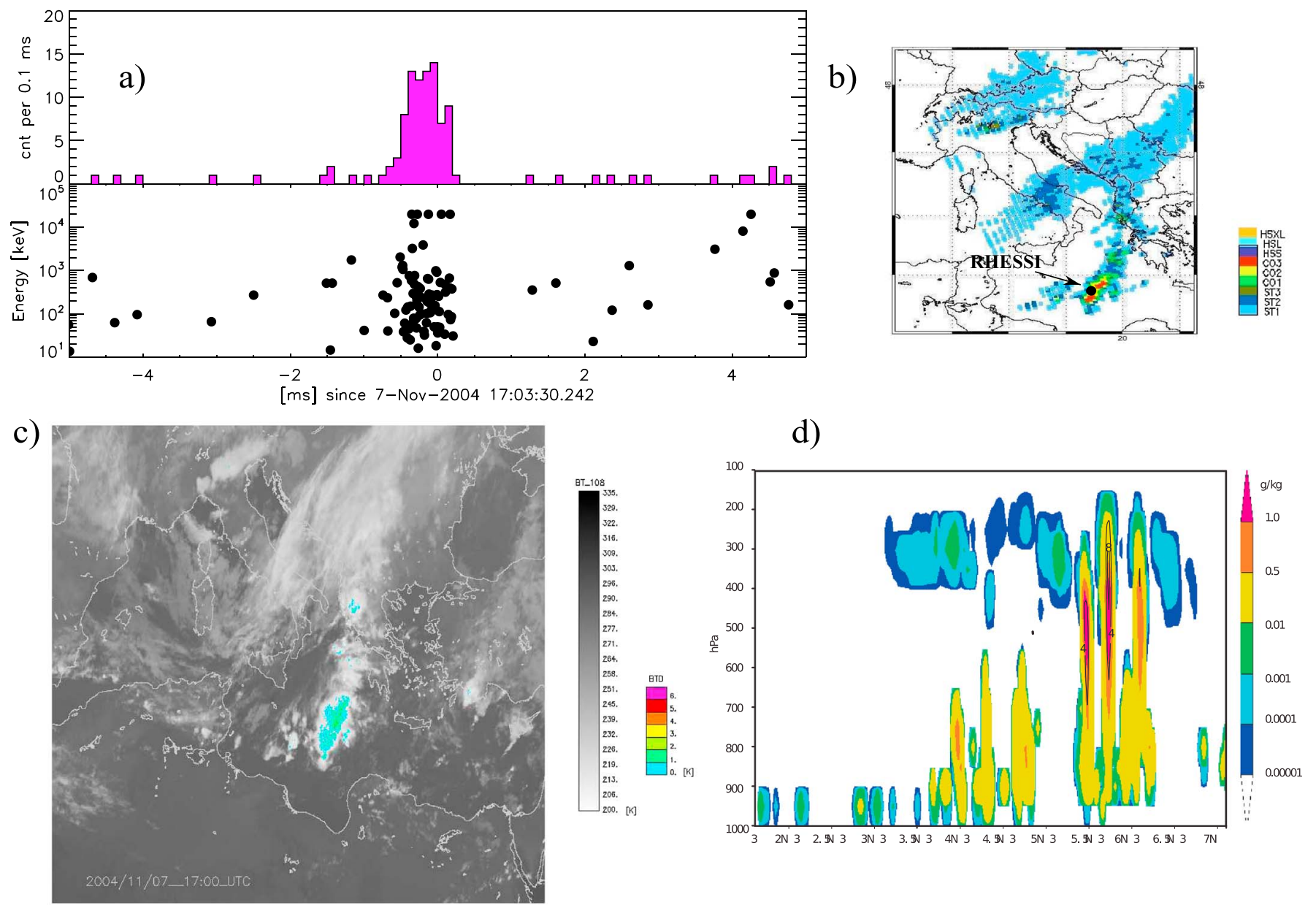

**Figure 2.** (a) The lightcurve and a scatterplot of photon energy and time for the RHESSI TGFs on 7 November 2004. (b) MWCC at 16:15 UT. Two deep convective cores (C03) are detected by the MWCC method correspondingly to the overshooting top retrieved by MSG. (c) MSG 10.8 μm cloud temperature measured at 17:00 in grey scale. BTD is shown in color. The maps range is latitude 25° to 50° and longitude −5° to 40°. (d) WRF model inner grid output: cloud water and cloud ice content (colors) and vertical velocity (contours) along the cross section where the convection is more developed: middle, cross section at 18.55°E, at 1500 UT. For the vertical velocity, only the values $w = 4, 8, 12$ m/s are shown.

The TGF was produced at the mature stage of the storm after a progressive increase in size for more than 7 h. At the moment of the TGF the storm reached an area of ∼ $23 \cdot 10^3$ km². Although from that moment the storm size kept stable for about 2 h, the convective part <−55°C started to increase in size for the next 2 h. At the time of the TGF the convective cores represented 34 % of the storm. At that moment the storm had the typical conditions for transient luminous event (TLE) production [*Soula et al.*, 2009].

Figure 3a shows the lightcurve and a scatterplot of the TGF measured by RHESSI. Here we have assumed that the WWLLN location was the source of the TGF. We have then corrected for the light travel time from source to RHESSI and also added the RHESSI offset of 1.8 ms [*Grefenstette et al.*, 2009; *Gjesteland et al.*, 2012]. This TGF contains two gamma pulses separated by 2.4 ms. The vertical lines show the times of lightning measured by the location networks. These times are also presented in Table 1.

WWLLN detected two strokes related to this TGF. The first WWLLN stroke occurred 0.6 ms before the first pulse of the TGF, and the second stroke occurred simultaneous with the second TGF pulse. Both WWLLN lightning had the same location 73 km east of the RHESSI nadir point.

LINET classified the stroke as a negative cloud to ground (−CG) with peak current of −81.6 kA. The location is <5 km away from the WWLLN stroke, and the time between the second WWLLN lightning and the LINET lightning is 48 μs. The lightning was measured outside of the LINET network, and it has an estimated uncertainty of 2.5 km (H. Betz, personal communication, 2015).

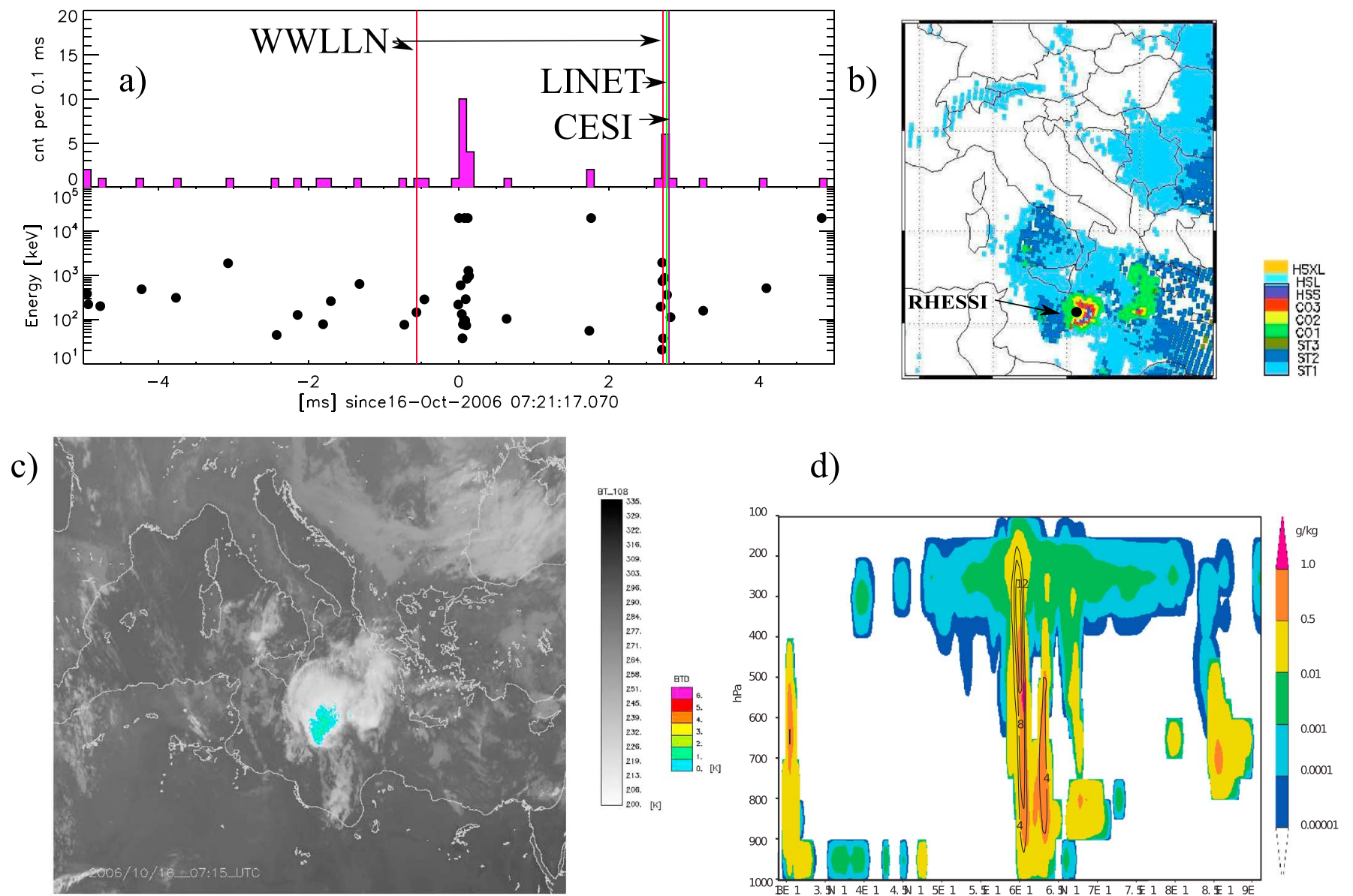

**Figure 3.** (a) The lightcurve and a scatterplot of photon energy and time for the RHESSI TGF on 16 October 2006. All RHESSI counts are time shifted back to the WWLLN source. The vertical lines indicate the time of the sferic measurements. WWLLN events are marked in red, the LINET is green, and CESI is purple. (b) MWCC at 04:54 UT. A compact deep convection with a well-defined overshooting top. The convection develops with a typical structure from weak convection to deep convection showing ice particles on top classified as hailstones. (c) MSG 10.8 μm cloud temperature measured at 07:15 in grey scale. BTD is shown in color. The maps range is latitude 25° to 50° and longitude −5° to 40°. (d) WRF model output cross section at 35.70°N, at 06:00 UT. For the vertical velocity, only the values $w = 4, 8, 12$ m/s are shown.

LINETs algorithm to distinguish between CG and IC works well when the distance between the lightning and the closest station is less than 125 km and the baseline between two stations are less than 250 km [*Betz et al.*, 2009]. The TGF occurred ∼700 km from the two closest sensors, and the baseline between the two sensors are ∼1000 km. When the LINET network started in 2006, the polarity of lightning detected far outside the network could be reversed. In 2007 this uncertainty was removed. The network can therefore not rule out the possibility that this flash is in fact of positive polarity and not negative (H. Betz, private communication, 2015).

The lightning location network operated by CESI classified the lightning as a cloud to ground (CG) stroke of −83.8 kA. The time of the stroke is ∼0.4 ms after the second pulse of the TGF, and it was located 128 km northwest of the WWLLN and LINET lightning. The sferic was located using two stations, being far out of the network domain territory. Therefore, it has a stretched uncertainty ellipse with major semiaxis of 99 km as shown in Figure 4. If the CESI source in fact was from the same location as the WWLLN sferics, it must occur earlier due to the light propagation time. If we use the speed of light as the velocity of the sferic propagating from the source to the receivers and move CESI to the end of the error ellipse (99 km), the CESI stroke must occur 0.33 ms earlier. This is the time used in Figure 3.

**Table 1.** Lightning Detection Related to the TGF on 16 October[a]

| Lightning Network | Latitude (°N) | Longitude (°E) | Time (s)[b] | Type |
|---|---|---|---|---|
| WWLLN (first stroke) | 35.3997 | 17.3051 | 17.069439 | NA |
| WWLLN (second stroke) | 35.4021 | 17.3307 | 17.072720 | NA |
| LINET | 35.4133 | 17.3443 | 17.072768 | −CG 81.6 kA |
| CESI | 36.0525 | 16.3592 | 17.073123 | −CG 83.8 kA |

[a]The source time of the two TGF pulses are 17.07028 and 17.07268. NA is for not applicable.
[b]Time is seconds since 16 October 2006 07:21:00.

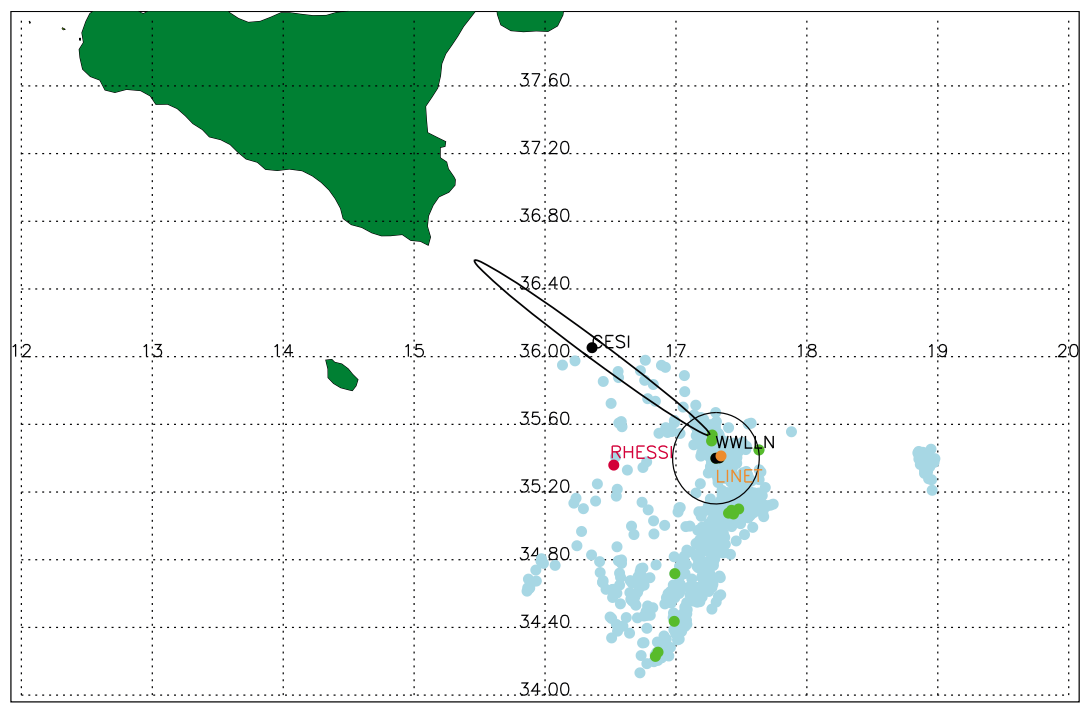

**Figure 4.** The map shows the RHESSI location for the TGF on 16 October 2004. The blue dots are lightning measured by WWLLN ±30 min at the time of the TGFs, while the green dots are WWLLN within ±30 s. The uncertainty ellipses of WWLLN and CESI is also shown. LINET uncertainty is 2.5 km.

Figure 4 shows a map of RHESSI subsatellite position and the lightning locations. Both WWLLN lightning strokes were located 73 km east of the RHESSI nadir point. The WWLLN 10 km uncertainty is also shown. The LINET lightning, which was located < 5 km from WWLLN is shown in orange.

The blue dots in Figure 4 are WWLLN lightning within ±30 min of the TGF, and the green dots are WWLLN lightning within ±30 s.

## 4. Discussion

### 4.1. The 27 May 2004 Event

This TGF occurred during spring. The convective cloud measured 54 min after the TGF is C01. This indicates convection limited altitude. The tropopause measured at Trapani station for this day was between 10 and 11 km altitude. This TGF did not come from a strong convective cloud. The MSG cloud top temperature measured at 20:30 UT indicates that the top of the cloud is located at 10 km altitude (316 g/cm$^2$).

### 4.2. The 7 November 2004 Event

The meteorological observations show high convection up to 8–10 km (414–316 g/cm$^2$) altitude and a cloud top at 12 km altitude (227 g/cm$^2$) 48 min before the time of the TGF. The cloud top temperature measured at 17:00 UT indicates that the top of the cloud was located at 12 km altitude (227 g/cm$^2$).

This TGF was so bright that it is possible to do a spectral analysis of the RHESSI TGF measurement. The TGF also occurred before the degeneration of the RHESSI detectors [*Grefenstette et al.*, 2009] such that the RHESSI detector response matrix can be used to compare simulated spectrum with the measurements.

Figure 5 shows the RHESSI energy spectrum with uncertainties in black. From our Monte Carlo model described in Østgaard et al. [2008] we simulate TGFs produced at 6, 8, 10, 12, and 14 km altitude. We use a typical Relativistic runaway electron avalanche (RREA) spectrum, $dN/dE \propto 1/E \times \exp(E/7.5\text{MeV})$. We also assume that RHESSI is within the TGF emission cone, which we assume to be 30° [*Gjesteland et al.*, 2011].

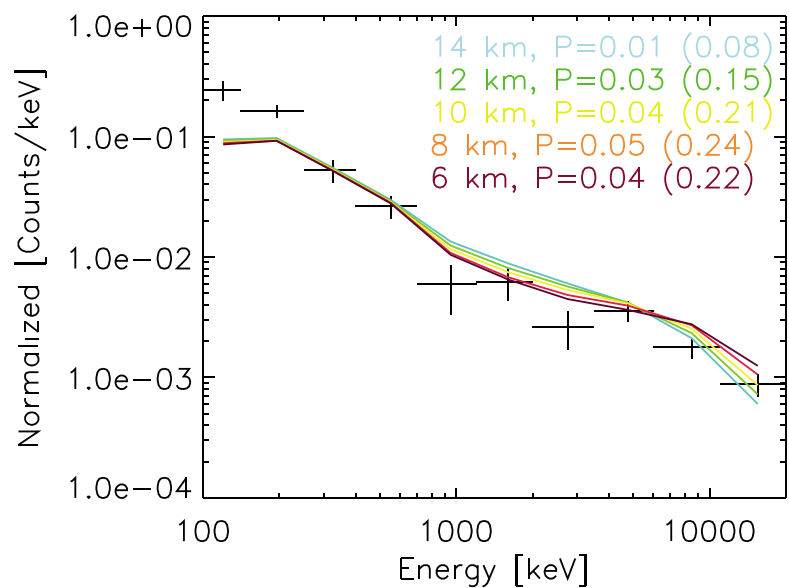

**Figure 5.** RHESSI TGF on 7 November 2004 energy spectrum with uncertainties is shown in black. The curves are the result from our MC simulation, with 6, 8, 10, 12, and 14 km production altitude, folded through the RHESSI detection response matrix and normalized to the RHESSI measurements. The $p$ value on the figure is the probability that the simulated curves fit the observations. The $p$ value in parentheses is calculated only using photons with energy >140 keV. A production altitude of 8 km gives the best fit to the measurements.

We have used the Pearsons chi-square test to compare our simulated spectra to the RHESSI measurements. The null hypothesis is that the simulated spectra and the measurements are equal. For each simulated spectrum we calculate the $p$ value as shown in Figure 5. The $p$ values shown in parentheses are calculated by using counts with energy > 140 keV (not including the first bin in Figure 5). The reason for also discussing the $p$ value without the 100–140 keV bin is that the aluminum cryostat that houses the detectors may influence the low-energy counts. We cannot reject the null hypothesis for simulated TGFs with production altitude 6–14 km (536–227 g/cm$^2$) at a significant level of 0.05. If we include the 100–140 keV bin, only 8 km (409 g/cm$^2$) production altitude has a $p$ value $\geq 0.05$.

From the meteorological analysis of the thundercloud 8–10 km (409–307 g/cm$^2$) corresponds to the top of the strong convection in the cloud. A TGF production altitude at this altitude is not rejected by the spectral analysis. The cloud top measured by MSG at 17:00 UT was 12 km (227 g/cm$^2$). If the TGF was produced at this altitude, it is also not rejected by the spectral analysis.

### 4.3. The 16 October 2006 Event

The meteorological observations show high convection up to 10 km (313 g/cm$^2$) altitude and a cloud top of $\geq 12$ km ($\leq 227$ g/cm$^2$) 2.5 h before the time of the TGF. The cloud top temperature measured at 7:15 UT indicates that the top of the cloud is located at 12 km altitude (234 g/cm$^2$). However, the WRF simulation shows that the cloud had towers that reached up to 150 g/cm$^2$.

For this TGF we have lightning location measurements from several networks. As shown in section 3.3, the second pulse of the TGF was simultaneous with lightning measured by three lightning location networks. All measurements are simultaneous to within 100 µs. However, the absolute timing of the RHESSI clock is under discussion. The RHESSI clock has been suggested to have 1–2 ms random uncertainty [*Grefenstette et al.*, 2009]. The time between the two pulses of this TGF is 2.4 ms. The time between the two WWLLN sferics is 3.3 ms. It is therefore not possible that both WWLLN sferics are caused by the two relativistic runaway avalanches that produced the two pulses of the TGF. It is likely to assume that the second TGF pulse caused the second WWLLN sferic and also possibly the LINET and CESI sferic. However, the first WWLLN sferic was not related to a gamma pulse that was measurable at satellite altitude.

If the measurements by WWLLN and LINET, which were simultaneous with the second TGF pulse, were caused by the RREA that also produced the TGF, this current must be of positive polarity, accelerating electrons upward. LINET classified this flash as being of negative polarity. However, LINET started in 2006 and at October 2006 the network was not fully calibrated. We can therefore not exclude the possibility that this stroke was, in fact, of positive polarity (H. Betz, personal communication). Also, in previous study the polarity of the current produced by RREA has been misclassified by other networks that automatically classify lightning.

**Table 2.** The Table Shows the Results From MWCC and MSG[a]

| TGF Time UT | MWCC | MSG Cloud Top Temperature |
|---|---|---|
| 27 May 2004 19:40 | C01 | 220–225 K |
|  | 8 km | 10 km |
|  | 535 g/cm$^2$ | 316 g/cm$^2$ |
|  | (measured 20:32 UT) | (measured 19:45 UT) |
| 7 Nov 2004 17:03 | C03 | 210–213 K |
|  | 10 km | 12 km |
|  | 307 g/cm$^2$ | 227 g/cm$^2$ |
|  | (measured 16:15 UT) | (measured 17:00) |
| 16 Oct 2006 07:21 | C03 | 205–210 K |
|  | 10 km | >12 km |
|  | 313 g/cm$^2$ | < 234 g/cm$^2$ |
|  | (measured 04:54 UT) | (measured 07:15 UT) |

[a]The time of the measurement is also given. The MSG cloud top temperature is the averaged value of the overshooting pixels (shown in color in Figures 1c, 2c, and 3c). The altitude estimation is calculated using a lapse rate of 6.5 K/km.

In a study by *Cummer et al.* [2013] sferics from National Lightning Detection Network (NLDN) were analyzed and they found that NLDN may get wrong polarity. They also found that NLDN in some cases classified an IC as a CG lightning. This happens especially for complex strokes such as the ones involving TGFs.

### 4.4. TGF Production Altitude

The observation of the convection and cloud top altitude for the thunderstorms are summarized in Table 2. Since the first analysis of the cumulative spectrum of RHESSI by *Dwyer and Smith* [2005], most studies have suggested that TGFs originate below the tropospause and are produced by electrical fields inside the thundercloud. *Williams et al.* [2006] estimated gamma attenuation and found that it is possible for TGF to be produced in the upper troposphere and make it to satellite altitude. Several studies have analyzed the TGF energy spectrum, and they all conclude that it is consistent with a production altitude ≤ 15 km [*Dwyer and Smith*, 2005; *Carlson et al.*, 2007; *Østgaard et al.*, 2008; *Gjesteland et al.*, 2010].

Detailed studies of the radio emission from TGFs have also shown that TGFs may be produced down to ~10 km altitude. For two TGFs *Stanley et al.* [2006] used ionospheric reflection to estimated the source altitude for the TGF-related sferics to be 13.6 km and 11.5 km. In a similar study by *Shao et al.* [2010] they found the TGF-related sferic origin to be between 10.5 and 14.1 km.

*Cummer et al.* [2014] reported two TGF measured by Fermi. These TGFs were estimated to be produced at 11.8 ± 0.4 km and 11.9 ± 0.94 km. This places the source region in the interior of the thunderstorm between the two main charge layers and implies an intrinsic TGF brightness of approximately 10$^{18}$ electrons.

*Lu et al.* [2010] reported on a RHESSI TGF which occurred between the negative and positive charge layers, which were located at 8.5 and 13 km altitude. In a time analysis of a TGF over Maracaibo lake by *Østgaard et al.* [2013] the TGF was found to occur close to the upper positive charge layer during the leader propagation of an IC lightning.

The most compelling theories for TGF production are the relativistic feedback discharge model [*Dwyer*, 2012] and cold RREA from lightning leaders as suggested by *Carlson et al.* [2009] and *Celestin and Pasko* [2011, 2012]. In both these theories the TGF must be produced between the midlevel negative and the main positive charge layer of the thundercloud.

Since the time between the MWCC convection measurements and the TGFs are 48 min, 54 min, and 2.5 h with respect to the TGF time, we cannot rule out that the TGF, in fact, are produced at higher altitude. However, since previous studies shows that TGFs are produced inside thunderclouds, we use the MSG cloud top estimate as an upper limit for TGF production.

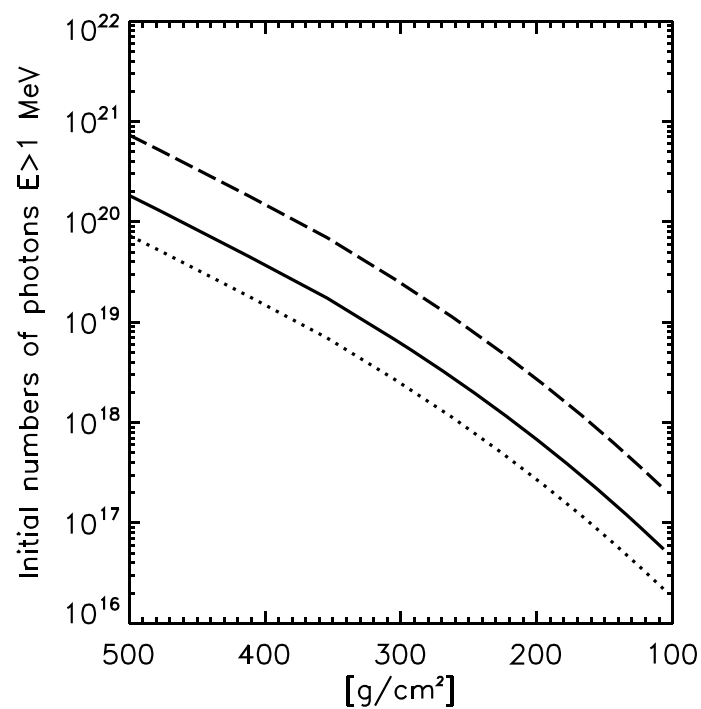

**Figure 6.** The curves show the estimated number of initial photons, with energy above 1 MeV, needed in our MC simulation as a TGF to escape atmospheric attenuation as a function of atmospheric column density. The various curves are for weak TGFs (∼10 photons measured by RHESSI), average (∼25 photons), and bright (∼100 photons). A column mass density of 500 g/cm$^2$ corresponds to approximate 6 km altitude and a column mass density of 100 g/cm$^2$ corresponds to approximate 17 km. The number of initial photons is approximately proportional to $\exp(\rho_{tot}/45 \text{ g/cm}^2)$ from 250–100 g/cm$^2$.

### 4.5. Estimation of Initial Number of Photons

For each TGF we estimate the number of initial photons with energy above 1 MeV, $N_{ph}(E > 1 \text{ MeV})$, which corresponds to the RHESSI measurements. This number is estimated by simulating TGF through the atmosphere to satellite altitude and comparing the simulated fluence to measurements. We can then scale the number of initial photons in the simulation to get an estimated initial number of photons. We have used the same simulation as presented in *Østgaard et al.* [2008] and *Gjesteland et al.* [2011] to propagate the photons from an initial altitude to the satellite. In the simulation we start with $5 \times 10^6$ photons with a typical RREA spectrum, $dN/dE \propto 1/E \times \exp(-E/7.5\text{MeV})$. The photon directions are uniform within a cone with half angle 30°.

Figure 3 in *Gjesteland et al.* [2011] shows the fluence as a function of nadir angle for TGF emission within a cone with half angles of 20°, 40°, and 60°. The scaling factor is similar as long as the TGF are measured inside the emission cone. If the TGF is measured outside the emission cone, only Compton scattered photons will reach the satellite and the scaling factor will increase. In such case, *Gjesteland et al.* [2011] found that the scaling factor increased by a factor of ∼4.

By using the effective area of RHESSI (239 cm$^2$, *Grefenstette et al.* [2008]) we can scale our simulation result to match RHESSI measurements. Figure 6 shows the initial number of photons needed in a TGF in order to escape atmospheric attenuation as a function of atmospheric mass. The dotted curve corresponds to weak TGFs (∼10 counts in RHESSI per TGF), the solid curve corresponds to average TGFs (∼25 counts) and the dashed curve corresponds to bright TGFs (∼100 counts). For the Mediterranean basin an altitude of 6 km corresponds to ∼500 g/cm$^2$ and 17 km altitude corresponds to ∼100 g/cm$^2$. These numbers are from the MSIS-E-90 atmosphere model [*Hedin*, 1991].

For the TGF 27 May 2004 and 7 November 2004 events we do not know the exact source location of the TGF. We therefore assume that the TGFs were produced at nadir. This introduces an error in the estimate of the number of initial photons in the TGF. If the TGF was produced off nadir, the initial number of photons must be increased to match the measured brightness. The fluence as a function of nadir angle is shown in Figure 3 in *Gjesteland et al.* [2011]. The variations in fluence are due to two factors. (1) The $R^2$ effect due to increasing distance. This scales as $(1/\cos(\alpha))^2$, where $\alpha$ is the angle between nadir and the TGF source location. (2) Increased attenuation due to increased column density. This scales as $\exp(1/\cos(\alpha))$. A TGF-measured 30° off nadir must therefore be 1.55 times brighter at the source than if it occurred at nadir. For RHESSI altitude (∼550 km) 30° corresponds to ∼300 km. From the meteorological data we see that the active part of the cloud, where we assume the TGF was produced, was closer than 300 km from nadir. Therefore, our estimate is conservative, and the number of photons at the source can be up to 1.6 times larger.

**Table 3.** Initial Number of Photons and Current Estimates for Most Likely Production Altitude[a]

| Date | Duration ($\sigma$)[b] | Altitude[c] | $N_{ph}(E > 1 \text{ MeV})$ | $I_{mom}$ | $\Delta M_q$ |
|---|---|---|---|---|---|
| 27 May 2004 | 78 ± 18μs | 6 km (535 g/cm$^2$) | 7 × 10$^{19}$ | 200×$\rho$ kA km | 39.1 × $\rho$ C km |
|  |  | 8 km (414 g/cm$^2$) | 2 × 10$^{19}$ | 90 × $\rho$ kA km | 17.7 × $\rho$ C km |
|  |  | 10 km (316 g/cm$^2$) | 3 × 10$^{18}$ | 21 × $\rho$ kA km | 4.2 × $\rho$ C km |
| 7 Nov 2004 | 204 ± 17μs | 6 km (536 g/cm$^2$) | 7 × 10$^{20}$ | 764×$\rho$ kA km | 390.9 × $\rho$ C km |
|  |  | 8 km (409 g/cm$^2$) | 2 × 10$^{20}$ | 345×$\rho$ kA km | 176.9 × $\rho$ C km |
|  |  | 10 km (307 g/cm$^2$) | 3 × 10$^{19}$ | 82×$\rho$ kA km | 42.0 × $\rho$ C km |
|  |  | 12 km (227 g/cm$^2$) | 6 × 10$^{18}$ | 26×$\rho$ kA km | 13.3 × $\rho$ C km |
| 16 Oct 2006 | 44 ± 9μs | 8 km (410 g/cm$^2$) | 3 × 10$^{18}$ | 160×$\rho$ kA km | 17.7 × $\rho$ C km |
|  |  | 10 km (313 g/cm$^2$) | 4 × 10$^{17}$ | 38×$\rho$ kA km | 4.2 × $\rho$ C km |
|  |  | 12 km (234 g/cm$^2$) | 4 × 10$^{17}$ | 10×$\rho$ kA km | 1.1 × $\rho$ C km |

[a]$\rho$ is the gamma production effective constant defined as $\rho = N_{re}/N_{ph}$, where $N_{re}$ and $N_{ph}$ are the number of electrons and photons both with energy above 1 MeV in a RREA.
[b]The duration given in this table is 1$\sigma$ of the gauss fit to the lightcurve.
[c]The atmospheric mass density is taken from MSIS-E-90 atmosphere model at the time and location of each TGF.

The TGFs on 27 May 2004 and 16 October 2006 were dim events which correspond to the dotted curve in Figure 6. The TGF on 7 November 2004 is among the brightest ever measured by RHESSI. For this TGF RHESSI measured 79 counts over an effective area of 239 cm$^2$. The initial brightness is shown as the dashed curve in Figure 6.

For the 27 May 2004 TGF the main convection was up to 8 km altitude with the cloud top at ∼10 km altitude. In the MSIS model 8 km altitude for this TGF corresponds to an atmospheric density of 409 g/cm$^2$ and 10 km corresponds to 307 g/cm$^2$. If we assume that this TGF occurred between the top of the main convection and the cloud top, it must contain between 3 × 10$^{18}$ and 2 × 10$^{19}$ initial photons with energy above 1 MeV. The TGF on 7 November 2004 is suggested to be produced between 10 km (307 g/cm$^2$) and 12 km (227 g/cm$^2$) altitude. Our estimate of initial photons for this TGF is between 6 × 10$^{18}$ and 3 × 10$^{19}$ photons. The TGF on 16 October 2006 had main convection up to 10 km (307 g/cm$^2$) with a cloud top at ≥12 km ( ≤227 g/cm$^2$). This TGF must then have between 4 × 10$^{17}$ and 3 × 10$^{18}$ initial photons. If the production altitude for this event was higher than 12 km (227 g/cm$^2$) it contained less than ∼ 10$^{17}$ photons. This is in the range of a typical RHESSI TGF. Our estimated initial number of photons is shown in Table 3.

Our estimates of initial photons for the event on 27 May 2004 and 7 November 2004 are larger than previous studies since we estimate a lower production altitude for the TGFs. *Dwyer and Smith* [2005] found that a typical RHESSI TGF contains 2 × 10$^{17}$ initial photons if they are produced at 15 km altitude. In their modeling 15 km altitude corresponds to 130 g/cm$^2$. In our simulation we find that 1.5 × 10$^{17}$ initial photons are needed for a TGF that propagates through 130 g/cm$^2$ atmosphere and results in 25 counts in RHESSI. This is shown by the solid line in Figure 6.

### 4.6. Current Moment and Charge Moment Estimate

Based on the number of initial photons in the TGFs, we can estimate the current moment for the TGFs. These current estimates are based on equation (24) from *Dwyer* [2012] and equation (1) from *Connaughton et al.* [2013] as $I_{mom}(0)$:

$$I_{mom}(t) = \frac{e\alpha\tau_a\mu_e E N_{re} \Delta z}{\sqrt{2\pi}\sigma} \exp\left(\frac{-t^2}{2\sigma^2}\right), \quad (1)$$

where $e$ is the electron charge (1.6 × 10$^{-19}$C), $\alpha$ is the ionization per unit length per runaway electron, $\mu_e$ is the mobility of the low-energy electrons, and $\tau_a$ is their attachment time. We have used values from *Morrow and Lowke* [1997] and at 10 km (307 g/cm$^2$) altitude $\alpha = 2645$ m$^{-1}$, $\mu_e = 0.26$ m$^2$/Vs, and $\tau_a = 5.17 \times 10^{-7}$ s. $E$ is the electrical field strength of the RREA region, which is assumed to be 285 kV/m at sea level and scaled with the inverse of the density [*Dwyer*, 2012]. $N_{re}$ is the total number of electrons with energy > 1 MeV, and $\Delta z$ is the average distance the runaway electrons travel and, according to *Dwyer* [2012], is 60 m at sea level and then scaled linearly with atmospheric density. The TGF lightcurve has been fitted with Gaussian distribution, and the $\sigma$ from this fit has been used in equation (1). For our estimates we have determined $\sigma$ from a maximum likelihood estimate of the RHESSI measurements. In our current estimate we let $N_{re}$ equal $\rho N_{ph}$, where $\rho$ is the number of photons with energy > 1 MeV produced per electron with energy > 1 MeV.

The result of our current moment estimates are presented in Table 3.

The ratio of electrons to photons varies with the electrical field strength of the RREA region. According to [*Skeltved et al.*, 2014], the ratio between photons and electrons (both with energy above 1 MeV) varies with the strength of the electrical field that produces the TGF. If the field is close to the runaway breakdown threshold (284 kV/m) the ratio is close to one to one. *Skeltved et al.* [2014] found the ratio ($N_{ph}/N_{re}$) to be 0.23 for an electrical field of 400 kV/m (scaled with air density), which corresponds to $\rho = 1/0.23 = 4.3$. They also show that $\rho$ is between 1 and 10 for electrical fields between 400 kV/m and 800 kV/m and between 10 and 100 for stronger electrical fields (800–2000 kV/m). *Dwyer* [2012] argues that most of the electrons are produced at the end of the avalanche where the electrical field is 285 kV/m (scaled with atmospheric density). In this case $\rho = 1$. However, if a significant portion of the electrons are produced at higher field strength $\rho$ may be larger. We must therefore consider this as order of magnitude estimate.

To estimate the total charge moment change, we need to integrate the current moment ($I_{mom}$) over time to get the charge moment ($\Delta M_q$).

$$\Delta M_q = \int_{-\infty}^{\infty} I_{mom}(t) dt = I_{mom}(0)\sigma\sqrt{2\pi}, \quad (2)$$

where $I_{mom}(t)$ and $\sigma$ are as in equation (1). The estimated current moment change is also presented in Table 3

Table 3 shows the peak current moment and total charge moment for the TGFs. If we assume that the TGF of 27 May 2004 was produced at 8 km (414 g/cm$^2$) altitude, we estimate that this TGF was produced together with a peak current moment of 90 × $\rho$ kA km a current moment change of 18 × $\rho$ C km. At 10 km (316 g/cm$^2$) we estimate a peak current moment of 21 × $\rho$ kA km a current moment change of 4 × $\rho$ C km.

For the TGF on 7 November 2004 we estimate a peak current moment change of 82 × $\rho$ kA km and current moment change of 42 × $\rho$ C km for 10 km (307 g/cm$^2$) production altitude and 26 × $\rho$ kA km and current moment change of 13 × $\rho$ C km for 12 km (227 g/cm$^2$) production altitude.

A study by *Cummer et al.* [2005] analyzed 13 sferic from TGFs. They found a current moment change for these discharges between 107 C km and 11 C km with 49 C km as the average value. *Lu et al.* [2011] studied 54 TGF-associated lightning signals and found charge moment changes between 10 C km and 200 C km with a median charge moment change of 64 C km.

For the bright TGF on 7 November 2004 our rough estimate of the charge moment change for a TGF produced at 10 km (307 g/cm$^2$) altitude is 42 × $\rho$ C km. For a TGF produced at 8 km (409 g/cm$^2$) altitude we estimate 176.8 × $\rho$ C km. This is higher than previously measured but within the same order of magnitude. The charge moment change estimate for 10 km (307 g/cm$^2$) production altitude or lower is consistent with previous measurements. Our values must be considered as order of magnitude estimates.

## 5. Summary

We have presented three TGFs that originated from clouds over the Mediterranean basin, which is a region of sparse TGF observations. The TGFs on 7 November 2004 and 16 October 2006 came from clouds with tops around 12 km altitude and from clouds that had strong convection and penetrated into the stratosphere. These clouds are typically associated with heavy precipitation and intense lightning activity. The event on 27 May 2004 came from clouds that had low convection and a cloud top at 10 km altitude.

All three cases represent common Mediterranean storms including MCS that are prolific producers of TLE. The fact that there is a lack of some special feature of the storms related to the exposed TGF might suggest that TGF are not so rare events but the low lightning activity at the studied midlatitudes diminishes the chance of detection.

Energy spectrum analysis of the 27 May 2004 TGF is consistent with a low production altitude between 6 and 12 km corresponding to 536–227 g/$cm^2$ atmospheric depth. The altitude range 8–10 km is in the upper part of the thundercloud. The estimated initial brightness of this TGF for 12 km production altitude is $6 \times 10^{18}$. Our estimate is 1 order of magnitude higher than that of an average RHESSI TGF suggesting that TGFs may have large fluctuations in intrinsic brightness.

Estimates of the current moment change related to the TGFs are consistent with previous measurements for production altitude of ≥8 km for the 27 May 2004 TGF and ≥10 km altitude for the 7 November 2004 and 16 October 2006.

By estimating the duration and initial number of photons in the TGFs we can give an order of magnitude estimate of the current moment and current moment change for these TGF. Our results are consistent with previous measurements.

**Acknowledgments**

This study was supported by the European Research Council under the European Union's Seventh Framework Programme (FP7/2007-2013)/ERC grant agreement 320839 and the Research Council of Norway under contracts 216872/F50 and 223252/F50 (CoE). We are grateful for the support of the European Science Foundation Research Network Project TEA-IS. We thank the RHESSI team for the use of RHESSI raw data and software and David Smith for his help with RHESSI data. We thank the institutions contributing to WWLLN (http://wwlln.net/). We thank Hans D. Betz and nowcast GmbH for the use of LINET data. We thank CESI-SIRF. We thank Oscar van der Velde for the discussions about the meteorological situation of the cases and Nicolau Pineda for the satellite image processing. This study was partially supported by research grants from the Spanish 167 Ministry of Economy and Competitiveness (MINECO)AYA2009-14027-C05-05, 168 AYA2011-29936-C05-04, and ESP2013-48032-C5-3-R and ESP2013-48032-C5-3-R. The data used for this publication are available upon request from the authors (thomas.gjesteland@uia.no).